\documentclass[twocolumn,showpacs,preprintnumbers,amsmath,amssymb]{revtex4}
\usepackage{graphicx}% Include figure files
\usepackage{dcolumn}% Align table columns on decimal point
\usepackage{bm}% bold math

\begin{document}

\preprint{APS/123-QED}

\title{Sound velocity and absorption measurements under high pressure using
picosecond ultrasonics in diamond anvil cell. Application to the
stability study of AlPdMn}

\author{F. Decremps$~^{1}$\cite{email}, L. Belliard$~^{2}$, B. Perrin$~^{2}$ and M. Gauthier$~^{1}$}

\affiliation{$~^{1}$Institut de Min\'{e}ralogie et Physique des
Milieux Condens\'{e}s, Universit\'{e} Pierre et Marie Curie-Paris 6,
CNRS UMR 7590, 140 rue de Lourmel, 75015 Paris, France.\\
$~^{2}$Institut des NanoSciences de Paris,
Universit\'{e} Pierre et Marie Curie-Paris 6, CNRS UMR 7588, 140 rue
de Lourmel, 75015 Paris, France.}

\date{\today}

\begin{abstract}
We report an innovative high pressure method combining the diamond
anvil cell device with the technique of picosecond ultrasonics. Such
an approach allows to accurately measure sound velocity and
attenuation of solids and liquids under pressure of tens of GPa,
overcoming all the drawbacks of traditional techniques. The power of
this new experimental technique is demonstrated in studies of
lattice dynamics, stability domain and relaxation process in a
metallic sample, a perfect single-grain AlPdMn quasicrystal, and
rare gas, neon and argon. Application to the study of defect-induced
lattice stability in AlPdMn up to 30 GPa is proposed. The present
work has potential for application in areas ranging from fundamental
problems in physics of solid and liquid state, which in turn could
be beneficial for various other scientific fields as Earth and
planetary science or material research.
\end{abstract}

% PACS, the Physics and Astronomy
                             % Classification Scheme.

%62.20.Dc Elasticity, elastic constants
%62.80.+f Ultrasonic relaxation (see also 43.35.Fj Ultrasonic relaxation processes in liquids and solids
%high pressure apparatus and techniques, see 07.35.+k
\maketitle

Acoustic properties of solids and liquids are very sensitive to
subtle changes in local or long-range order which thus make
experimental measurements of phonons velocity and absorption under
high pressure one of the most useful probe of interatomic potentials
variations. In addition to provide a crucial test for modern ab
initio calculations or Earth and planetary models, elasticity under extreme conditions substantially
contributes to the understanding of phase transition mechanisms or
structural stability. In a same manner, attenuation measurements provide
useful information on relaxation processes for understanding the correlation
between acoustic waves and intrinsic excitations. However, to our knowledge no acoustic
attenuation as a function of pressure has been published and only
few specialized institutions have the capability of measuring in
laboratory sound velocity under high pressure and high or low
temperature. Experimental methods could be classified into two types: (i) the conventional pulse-echo ultrasonic
technique (acoustic wave generated by a piezoelectric transducer)
combined with a large-volume cell able to accommodate sufficiently large samples so that successive echoes do not
overlap~\cite{gauthier, spetzler}, and (ii) the Brillouin scattering
in diamond anvil cell (DAC)~\cite{polian}. The price to pay for the
first method is the centimeter dimensions of the experimental
volume, more than 4 orders of magnitude over the capabilities of DAC
devices, and for the second one the transparency of the sample. In
spite of many efforts to improve the pressure range, both
constraints still preclude studying the elastic properties in
laboratory of opaque materials under pressure above 10 GPa. In this
Letter, we report a novel approach, the picosecond ultrasonics technique in
diamond anvil cell, which can be easily set-up in laboratory and
circumvents all previous limitations\cite{vallee}. The main other
advantages can be summarized as follows. It covers a large frequency
range (from 1 GHz to 300 GHz) that bridges the frequency gap
between the usual laboratory techniques (from 1 MHz to 30 GHz with
ultrasonics and Brillouin scattering) and the large instruments one
(from 1 to 10 THz for inelastic neutron or X-ray scattering). It
opens a way of determining acoustic attenuation pressure dependence.
And, last but not least, it can be applied to the study of any
materials.

Time-resolved picosecond optical measurement in DAC will be first
described. The capability of this method will then be demonstrated
through the generation and detection of longitudinal hypersonic
sound waves in a single-quasicrystalline icosahedral AlPdMn up to
30 GPa using Ne or Ar as pressure transmitting medium (PTM). Several
reasons justify to select the ternary alloy
Al$_{68.2}$Pd$_{22.8}$Mn$_{9}$ for this study. It is one of the
metallic sample which can be routinely obtained with a high degree of
structural perfection and be prepared as thin platelets of excellent
polishing quality. As a consequence, its acoustic
properties are free from grain diffraction or macroscopic defects
such as surface roughness. The second reason is related to the
crucial role that play the acoustic data in the on-going debate of
the entropy contribution to the quasicrystals (QCs) stability\cite{henley}.
Experimentally, QCs have been found to have a local icosahedral
symmetry with a spatial coherence length close to that of the
silicon standards despite the absence of translational symmetry.
Nonpropagating local disorder as froze tiles flip\cite{coddens},
tunneling states\cite{bert} or the activation of a new type of
dynamics due to atomic jumps (called phasons\cite{henley}) are some
of the most interesting and still not yet understood consequences.
Within this context, measurements of sound velocity and attenuation
in perfect icosahedral AlPdMn as a function of pressure (\emph{i.e.}
density) may provide insight into the thermodynamic stability of
QCs.

\begin{figure}
\includegraphics[width=8.7 cm]{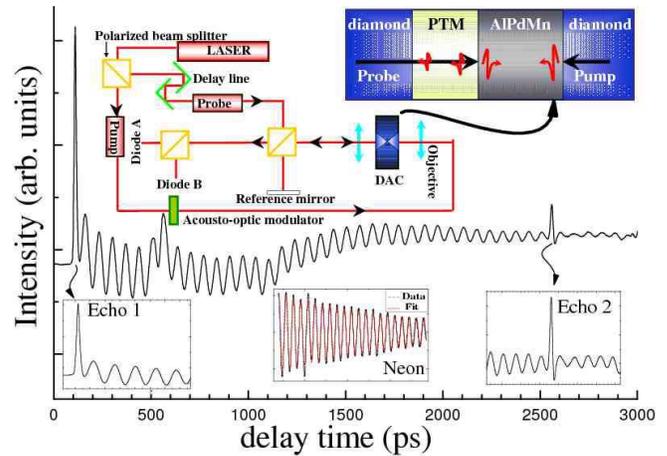}
\caption{\label{signal} Change in the reflectivity imaginary part of AlPdMn
at 9.7 GPa as a function of the optical probe-pulse time delay. Up:
experimental apparatus used to perform picosecond-laser acoustics
studies at high pressure. The arrow indicates the schematic
illustration of generation and detection process for AlPdMn in DAC.
Insets : left: enlarged part of the first acoustic echo. Center:
Brillouin interferences between the probe beam and the acoustic wave
in the PTM where the thermal background has been subtracted. The red
line corresponds to the data fit carried out to determine the
attenuation. Right: enlarged part of the second acoustic echo.}
\end{figure}

Nondestructive picosecond acoustic experiments belong to the family
of modern ultrafast optical methods used for about twenty years,
mainly carried out to measure the mechanical and thermal properties
of thin-film materials\cite{maris, antonelli}. A short laser pulse is split into pump and probe beams.
The pump is focused on an optically absorbing film surface, whereas the probe is focused to a spot of 3 $\mu$m on the opposite one.
The pump laser pulse creates a sudden and small
temperature rise (1 K). The corresponding thermal stress generated
by thermal expansion relaxes by launching an acoustic strain field.
After propagation along the sample, both thermal and acoustic
effects alter the optical reflectivity of the sample in two ways: the photo-elastic effect,
and the surface displacement as
the acoustic echo reaches the surface. The first modification contributes to
the change of both real and imaginary parts of the reflectance, whereas the
second one only modifies the imaginary contribution. The variation
of reflectivity as a function of time is detected through the
intensity modification of the probe delayed from the pump with a
different optical path length. In this study, we have adapted the
optical pump-probe set-up reported in previous
publications\cite{belliard} in such a way that measurements of
acoustic echoes in DAC be possible. Ultrashort pulses (800 nm, 100
fs) are generated by a Ti:Sapphire laser. The power in
each pump and probe pulse applied to the sample are about 20 mW and
3 mW respectively. The detection is carried out by a
stabilized Michelson interferometer which allows the determination
of the reflectivity imaginary part change\cite{perrin1}. Extracted from a large single
quasicrystal of icosahedral Al$_{68.2}$Pd$_{22.8}$Mn$_{9}$
($\rho$=5.08(1) g.cm$^{-3}$)~\cite{DeBoissieu} previously used to
determine accurate equation of state from both ultrasonics and X-ray
diffraction techniques\cite{decremps}, a thin platelet was loaded
into the experimental volume of a DAC. Two
experimental runs have been carried out with sample embedded in
different PTM~\cite{Couzinet}, neon and argon (see Fig.~\ref{signal}). Surprisingly, we systematically observed
a small amount of rare gas (thickness of about 1 $\mu$m) between the anvil and the pump-side AlPdMn surface.
Using the well known acoustic velocity of AlPdMn at ambient conditions\cite{decremps}, the
thickness of the single-grain was measured to be 9.12(1) $\mu$m from
picosecond measurement. Pressure was measured using the fluorescence
emission of a ruby sphere~\cite{Mao} placed close to the sample in
the gasket hole. The accuracy was better than 0.3 GPa at the maximum
pressure reached.

\begin{figure}
\includegraphics[width=8.7 cm]{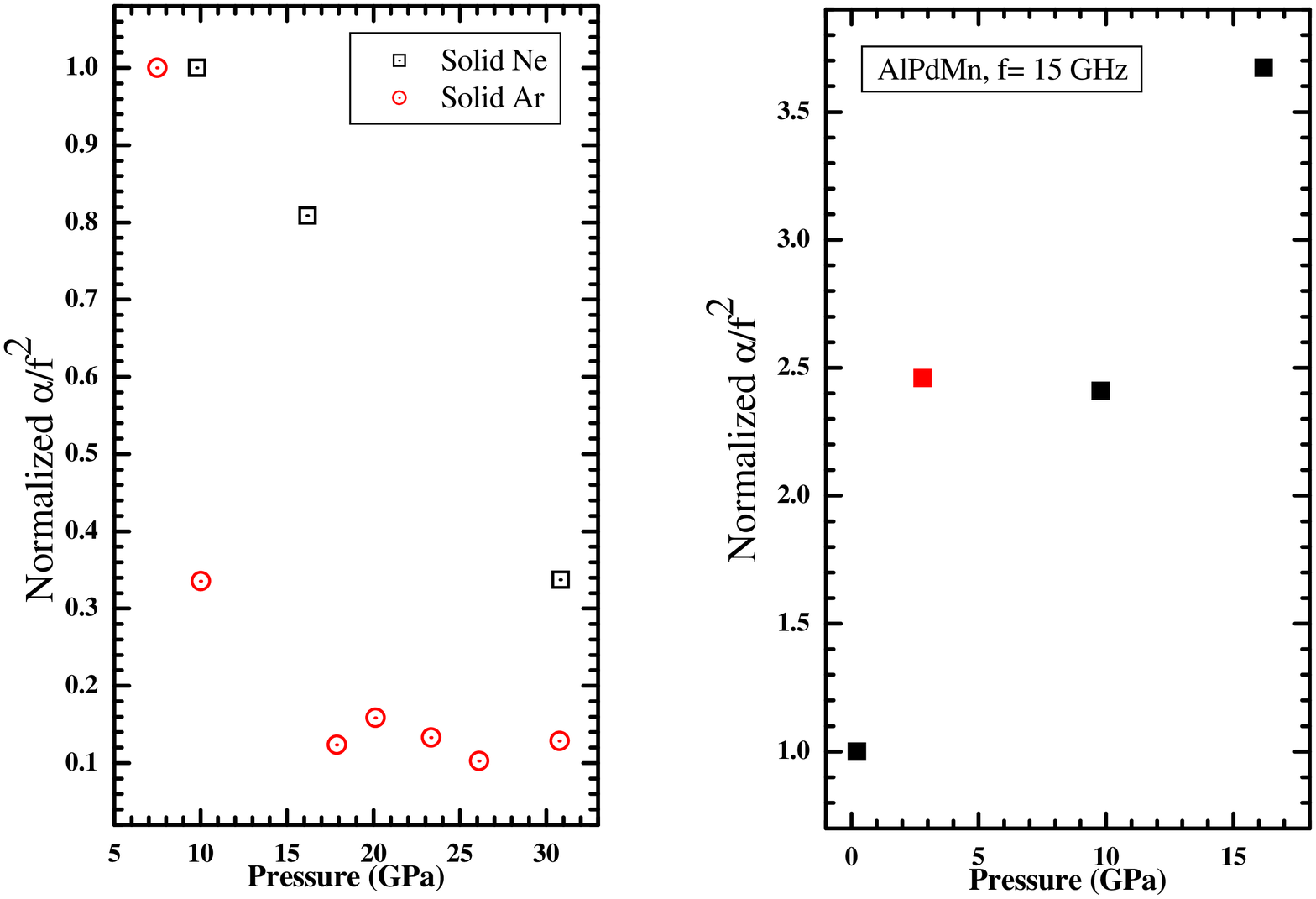}
\caption{\label{attenuation} Pressure dependence of the longitudinal
attenuation coefficient divided by the square of frequency and
normalized by the first measured point (P$_i$). Left: neon (P$_i$ =
7.5 GPa) and argon (P$_i$ = 9.7 GPa). Right: AlPdMn (P$_i$ = 0.2 GPa),
the black (red) full square correspond to the upstroke
(downstroke) data.}
\end{figure}

At each pressure, two longitudinal acoustic echoes were
systematically observed in the relative variation of the
reflectance imaginary part. The first echo corresponds to a single way between the
two surfaces of the sample, the second echo to a pulse being twice
reflected back  at the AlPdMn/PTM interfaces. As can be
seen in Fig.~\ref{signal}, the acoustic signal is superimposed on a
slow background variation caused by thermal effect plus a damped
oscillatory component. This last contribution arises from the
acousto-optic modulation of the probe beam from the acoustic waves
propagating in the PTM. It has thus been treated as usually done for
stimulated Brillouin scattering signals~\cite{maris} in order to
extract both velocity and attenuation in the PTM (neon or
argon). We obtain a pressure dependence of the sound velocity in
argon and neon in excellent agreement with previously reported
data\cite{shimizu}. Results for $\alpha / f^2$, the longitudinal
attenuation coefficient divided by the square of frequency, as a
function of pressure are shown in Fig.~\ref{attenuation}. We observe a
negative slope of the absorption vs density which, in the case of
rare gases, is related to the pressure dependence of the interaction between thermal
phonons and acoustic waves.

\begin{figure}
\includegraphics[width=8.7 cm]{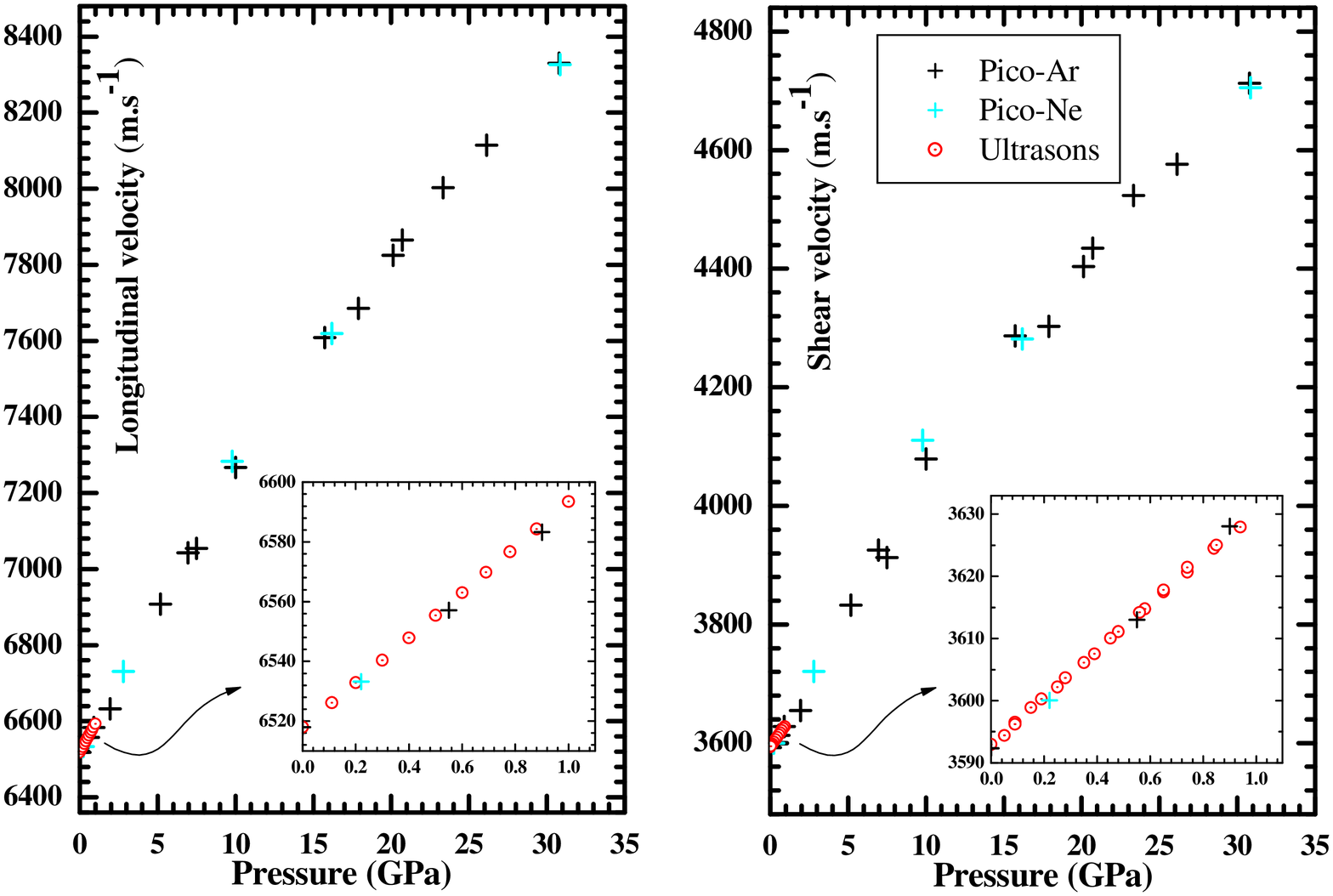}
\caption{\label{velocity} Left: longitudinal sound velocity of
AlPdMn as a function of pressure. Right: shear sound velocity of
AlPdMn as a function of pressure. Red circles: ultrasonic
data\cite{decremps}. Black crosses: picosecond data from the first
run using argon as PTM. Cyan crosses: picosecond data from the
second run using neon as PTM. In both runs, no difference between
upstroke and downstroke data was detected.}
\end{figure}

For AlPdMn, the longitudinal velocity $v_L$ has been determined using
the following equation: $v_L=2d/\Delta t$ , where $\Delta
t$ is the time interval between the two adjacent acoustic echoes
and $d$ the thickness of the sample deduced at high pressure from
its initial value and the pressure dependence of the
bulk modulus $B$~\cite{decremps}. The transversal
sound velocity $v_T$ is finally extracted from the same equation of
state and the definition of the bulk modulus for acoustically
isotropic materials: $B=\rho(v_L^2-4/3v_T^2)$. Pressure dependence of longitudinal and shear waves
is displayed in Fig.~\ref{velocity}. Here again, the excellent
agreement with previous ultrasonic study under hydrostatic pressure
up to 1 GPa demonstrates the capability of the present method. Using
these results, and according to the large pressure range probed as
well as the high accuracy of our technique (the travel time is
determined with an error of less than 2 ps corresponding to an
uncertainty on the velocity lower than 0.5 \%) we propose the
following non-linear curve fit of the elastic moduli (in GPa) versus pressure (in GPa): $C_{11}=215.88+7.33P-0.024P^2$, and
$C_{44}=65.6+2.53P-0.010P^2$. These results demonstrate that fourth
and higher order elastic moduli, known to play a major role in the
vicinity of phase transition, can be readily determined using our
technique.

The attenuation per unit length in AlPdMn is computed from the
ratio of the first and second echo Fourier amplitude
(respectively $\Delta R_1(\omega)$ and $\Delta
R_2(\omega)$)~\cite{morah}: $\alpha (\omega)=(2d)^{-1}ln|\Delta
R_1(\omega)*r^2_{QP} / \Delta R_2(\omega)|$ with $r_{QP}$ the amplitude
reflection coefficients for the phonon strain pulse at the interface between the
quasicrystal and the PTM.
This coefficient is dependent on the acoustic
impedances $\rho v$ only and has been determined under pressure using
Ref.~\cite{reflection, shimizu}. The pressure dependence of the
longitudinal attenuation coefficient divided by the frequency squared
(15 GHz) is compared to the rare gas results in
Fig.~\ref{attenuation}.

\begin{figure}
\includegraphics[width=6 cm]{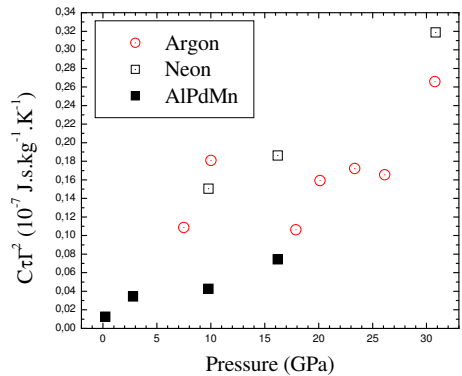}
\caption{\label{akhiezer} Pressure dependence of the
$C\tau{\Gamma}^2$ Akhieser term.}
\end{figure}

An obvious dissimilarity between the pressure behavior of
$\alpha$(P) in solid Ne (or Ar) and AlPdMn is observed. An
explanation to account for such a peculiar result could be related
to the interaction between phonons and dynamical defects in QCs:
increasing temperature above a few hundreds degrees C is known to
allow atomic jumps corresponding to the creation of structural
defects that have no counterpart in periodic solids. The activation
of this phason dynamics in QCs has been found to be at the origin of
phase transition\cite{coddens2} or strong deviation from regular
crystal behavior\cite{rochal}. However at 300 K the phasons are
undoubtedly frozen and the phonon-phason coupling hypothesis has
here to be abandoned. On the other hand, the existence of small
phonon-like atomic displacements due to an initial static disorder
of freeze tiles flip\cite{coddens} and/or the occurring of intrinsic
tunneling states\cite{bert} have been invoked to be mainly
\emph{responsible} for the stability of the quasicrystalline phase
at low temperature. Both effects are different from long-range
atomic jump and are expected to be increased by lattice contraction.
These scenarios could be an explanation for the singular behavior of
pressure-induced sound absorbtion through the existence of a
secondary elastic strain due to short-range atomic displacement (the
freeze tiles flip hypothesis) or of a tunneling state-phonon
coupling. The origin of relaxation in QCs at temperature lower than
the phason dynamics activation is thus quite different than regular
crystals, but the intrinsic mechanism should still be analytically represented
through the Akhieser formalism\cite{landau}:

\begin{eqnarray}
\frac{\alpha}{f^2} = \frac{2 {\pi}^2 T C \tau}{{v_L}^3}
{\Gamma}^2%
\label{eq:one}
\end{eqnarray}

where $C$ is the specific heat per unit mass of the crystal, $T$
the temperature, $\tau$ the mean lifetime of thermal excitations, and ${\Gamma}^2$ the visco-elastic contribution to the
sound absorption. The pressure
variation of $C \tau {\Gamma}^2$ term in AlPdMn, shown in
Fig.~\ref{akhiezer}, follows the same
behavior as those from rare gases despite the difference on the
attenuation. This is in agreement with
Duquesne\cite{duquesne} who has demonstrated that the thermoelastic
contribution to the process of sound absorption in AlPdMn is negligible in the
GHz range with respect to the viscosity of the phonons gas (100 K $<$ T $<$ 300 K). Lattice contraction effect
on the sound absorption could thus be interpreted through a peculiar Akhieser
contribution due to interaction between acoustic waves and short-wavelength fluctuations.
These intrinsic excitations could be invoked into the extraordinary stability of QCs versus
pressure (the pressure transition of AlPdMn is higher than 80
GPa\cite{hasegawa} and still unknown), as well as the
pressure-induced phase transition from an approximant to a
quasicrystal\cite{watanuki}. However, we point out that our interpretation should
be considered as a tentative which is now addressed
theoretically in order to get microscopic insight into the
relaxation phenomena in phason-free QCs. This would \emph{in fine}
provide information on the quasiperiodic long-range order
propagation and on the quasicrystal stability.

To conclude, we have demonstrated through the study of AlPdMn the
feasibility of carrying out picosecond acoustics combined with the
most versatile and powerful high pressure device, the DAC. In both
cases, a small dimension of the sample is mandatory. This
new high pressure tool makes possible accurate measurements of
attenuation (not amenable with any other traditional techniques),
and velocity of longitudinal waves in the tens of GHz range. We
measured a singular behavior of pressure-induced sound attenuation in
AlPdMn, which gives new experimental insight into the intrinsic
effect that stabilizes the free of long-wavelength phason fluctuations quasicrystalline phase.
This method can be easily extended to elastic investigations of all
materials (opaque, transparent, single- or polycrystal,
nanomaterials, or liquids) up to several Mbar and thousands of K. We
believe that the technique of picosecond acoustics in DAC is a
critical step forward to the study of elastic properties under
extremes conditions. Finally, it has to be pointed out that recent
works indicates a conceivable extension of the present development
to the detection and generation of shear waves\cite{rossignol}, as
well as the investigation of thermal properties\cite{perrin} as a
function of pressure and temperature. So far, this technique is
likely to have an impact on the study of dynamics of solid and
liquid states at high density.

%\begin{acknowledgments}
We thank G. Coddens for fruitful discussions and A. Polian for his critical reading
of the manuscript. We also wish to thank M. de Boissieu, providing the \emph{i}-AlPdMn
single-grain.
%\end{acknowledgments}

\end{document}